# Estimating Surface Sediments Using Multibeam Sonar

*Acoustic Backscatter Processing for Characterization and Mapping of the Ocean Bottom*

**By Frank W. Bentrem**
*Physicist*
and
**William E. Avera**
*Geophysicist*
and
**John Sample**
*Computer Scientist*
*Naval Research Laboratory*
*Stennis Space Center, Mississippi*

Knowledge of the seafloor sediments in littoral regions is of interest to such diverse groups as the world's navies, fisheries, marine construction companies and the U.S. Environmental Protection Agency (EPA). Navies are concerned with sonar performance, mine warfare and antisubmarine operations. Fisheries search for marine-life habitats, while construction companies need to plan routes for laying pipelines and cables. Finally, the EPA wishes to track absorption of contaminants, and the absorption rate is affected by the sediment type.

Traditional methods for obtaining information on seafloor sediments are slow and labor intensive. Divers may take underwater photographs, and survey vessels often procure grab samples and cores. The photos are then analyzed, and the grab samples and cores are taken to a laboratory to determine sediment-grain size distributions. Acoustic remote sensing of the seafloor advanced greatly with the advent of multibeam sonar. Though primarily used for obtaining bathymetry, the wealth of data available makes the goal of bottom characterization from multibeam systems enticing. This article describes an acoustic remote sensing technique that was developed (and packaged as SediMap®) in the Marine Geosciences Division at the U.S. Naval Research Laboratory.[1] The article also discusses the impact of this technology on oceanographic surveys and geologic interpretation.

## Model and Technique

Multibeam sonars receive sound beams from multiple angles along the cross-track swath. The acoustic intensity corresponding to the echo from each beam may be recorded for every ping. The strength of the seafloor scattering in the direction of the sonar is called "backscattering," strength and is derived from the echo intensities. This results in data that has backscattering strength as a function of beam angle.

This backscattering strength data may be compared to the acoustic backscattering model derived by Darrel Jackson and others at the Applied Physics Laboratory at the University of Washington (APL-UW).[2] The model predicts backscattering strength as a function of the seafloor-incidence angle. By tracing the sound rays in the water, incidence angles may be converted to beam angles. Among the six input values for the APL-UW model are sediment density, sediment sound speed and roughness parameters. Based on physical considerations and emperical data, these input parameters are mapped to mean-grain size.

A least-squares best fit for the acoustic model with the backscattering strength versus beam angle data[3] is found using the simulated annealing optimization technique.[4] The best-fit parameters are then mapped to mean grain size for an estimate of the bottom sediment. Averaging over groups of pings, these steps are repeated until sediment estimates are obtained for the entire survey.

## Multibeam Systems

SediMap has been tested using Kongsberg Maritime (Kongsberg, Norway) EM 121 and EM 1002 multibeam echo sounders. The EM

121 echo sounder operates at a frequency of 12 kilohertz and forms 121 beams across the ship track in a 120° swath with 1° separation among beams. Good results may be obtained in water depths of 100 to 400 meters. For shallow areas (10 to 200 meters), the EM 1002 echo sounder is appropriate. The EM 1002 operates at 95 kilohertz with 111 beams in a swath up to 150°. It should be noted that the echo sounders report accurate bathymetry for much larger water depth ranges. The ranges given above indicate the water depths for which the authors of this article have tested and gained confidence in the recorded backscattering strengths.

## Results

Bottom sediments in the testing areas ranged from clay to gravel in water depths from 10 to 400 meters. SediMap estimates from three surveys were compared to 38 core and grab samples with a grain size correlation 0.72. A sea trial onboard a Naval Oceanographic Office T-AGS 60 survey vessel effectively demonstrated the utility of using SediMap for the planning of core locations. The backscattering data was processed in near-real time, and patches of coarse sediments were discovered and verified by obtaining cores. In fact, the coarse sediment patches could not be distinguished from the fine sediment by viewing the multibeam imagery, but was clearly identified with the SediMap results.

## Enhancement with Segmentation

If the survey area has dense coverage (e.g., multiple parallel tracks with some overlap in the swaths), the data may be grouped by areas of similar acoustic "texture". This is accomplished with an image segmentation algorithm. The algorithm compares the histograms of every region in the acoustic image and uses a maximum likelihood method to separate the image into segments of similar texture.[5] Data from each image segment is then averaged and processed to obtain sediment grain size estimates in those segments. After image segmentation, far fewer sediment estimates are required, resulting in a greatly reduced processing time.

## Impact for Oceanography

SediMap will be transitioned to the Naval Oceanographic Office and is likely to be utilized as a tool for efficient planning for core and grab sample locations, as a guide for geologists in mapping seafloor sediments and to be used in the absence of ground truth, as seafloor estimates for acoustic modeling and archiving.

## Acknowledgements

The authors would like to thank Peggy Haeger and Gene Kelly of the Naval Oceanographic Office for providing multibeam data and ground truth. We also acknowledge Kevyn Malpass for his assistance with computer coding. The authors would also like to acknowledge sponsorship under Program Element 0603704N by the Oceanographer of the Navy via Space and Naval Warfare Systems Program Management Warfare 180, Capt. Wang, program manager. */st/*

For more information on this subject matter, visit the Web site at www.sea-technology.com and click on the title of this article in the Table of Contents.

*Frank W. Bentrem is a physicist in the Marine Geosciences Division at the Naval Research Laboratory where his research efforts have focused on the inversion of high-frequency acoustic backscatter. He received both an M.S. in physics and a Ph.D. in scientific computing from The University of Southern Mississippi.*

*William E. Avera is a geophysicist with 24 years of experience in shallow Earth and near-surface geophysics. He has worked in the Mapping, Charting and Geodesy branch of the Naval Research Laboratory since 1986. Prior to that position, Avera worked in reflection seismology for four years in the oil industry for ARCO Exploration Co. conducting oil and gas exploration in the Gulf of Mexico region.*

*John Sample has been a computer scientist with the Mapping, Charting and Geodesy branch since 1998, and is currently the technical lead for Naval Research Laboratory's Geospatial Information Database Portal. Sample has authored two book chapters and several conference proceedings related to geospatial and environmental data dissemination.*

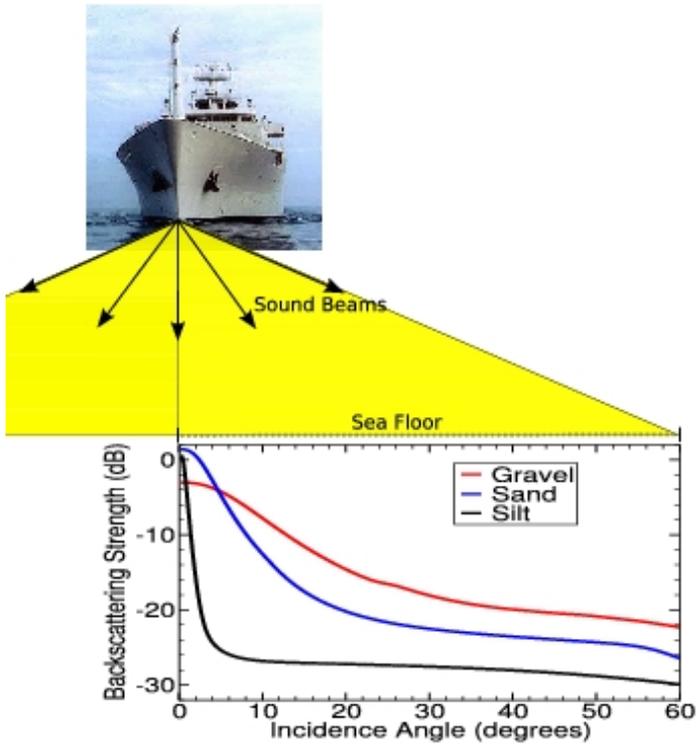

*Hull-mounted multibeam sonar shown with sound beams intersecting the seafloor at multiple incidence angles. The backscattering strength is a function of both the incidence angle and seafloor properties.*

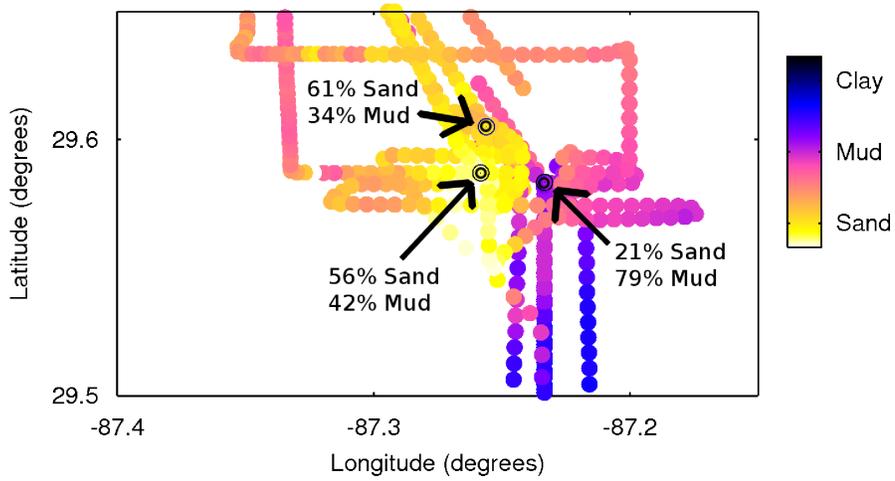

*Sediment grain size estimates in the Gulf of Mexico compared with core-sample analysis. Though sandy and muddy areas were visually indistinguishable in the acoustic imagery, SediMap® correctly identified the two sediment regions.*

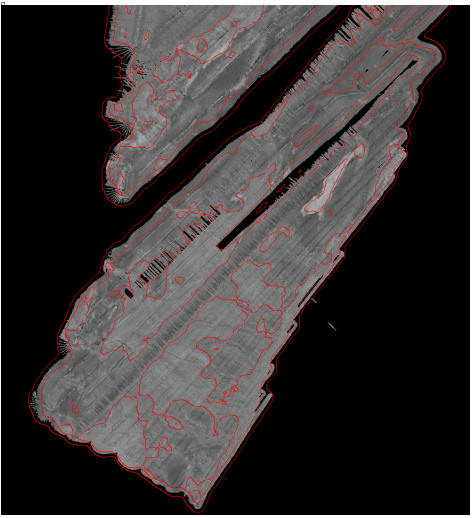

*Segmentation of acoustic imagery into similar textures--- few sediment estimates are required resulting in a much shorter processing time.*